\documentclass[twocolumn,english,aps,prl,showpacs]{revtex4}
\usepackage[T1]{fontenc}
\usepackage[latin1]{inputenc}
\usepackage{amsmath}
\usepackage{graphicx}
\usepackage{amssymb}
\usepackage{esint}

\makeatletter
\@ifundefined{textcolor}{}
{%
 \definecolor{BLACK}{gray}{0}
 \definecolor{WHITE}{gray}{1}
 \definecolor{RED}{rgb}{1,0,0}
 \definecolor{GREEN}{rgb}{0,1,0}
 \definecolor{BLUE}{rgb}{0,0,1}
 \definecolor{CYAN}{cmyk}{1,0,0,0}
 \definecolor{MAGENTA}{cmyk}{0,1,0,0}
 \definecolor{YELLOW}{cmyk}{0,0,1,0}
 }

\@ifundefined{definecolor}{\@ifundefined{definecolor}{\@ifundefined{definecolor}
 {\usepackage{color}}{}
}{}}{}\makeatother

\makeatother

\usepackage{babel}

\makeatother

\usepackage{babel}

\makeatother

\usepackage{babel}

\makeatother

\usepackage{babel}

\makeatother

\usepackage{babel}

\begin{document}

\title{Spin-orbit coupled weakly interacting Bose-Einstein condensates in
harmonic traps}

\author{Hui Hu$^{1}$, B. Ramachandhran$^{2}$, Han Pu$^{2}$, and Xia-Ji
Liu$^{1}$}

\affiliation{$^{1}$ACQAO and Centre for Atom Optics and Ultrafast Spectroscopy,
Swinburne University of Technology, Melbourne 3122, Australia \\
 $^{2}$Department of Physics and Astronomy, and Rice Quantum Institute,
Rice University, Houston, TX 77251, USA}

\date{\today}
\begin{abstract}
We investigate theoretically the phase diagram of a spin-orbit coupled
Bose gas in two-dimensional harmonic traps. We show that at strong 
spin-orbit coupling the single-particle spectrum decomposes into different 
manifolds separated by $\hbar \omega_{\perp}$, where $\omega_{\perp}$
is the trapping frequency. For a weakly
interacting gas, quantum states with skyrmion lattice patterns emerge
spontaneously and preserve either parity symmetry or combined parity-time-reversal
symmetry. These phases can be readily observed in a spin-orbit coupled gas of $^{87}$Rb
atoms in a highly oblate trap. 
\end{abstract}


\maketitle
Spin-orbit (SO) coupling leads to many fundamental phenomena in a
wide range of quantum systems from nuclear physics, condensed matter
physics to atomic physics. For instance, in electronic condensed matter
systems SO coupling can lead to quantum spin Hall states or
topological insulators \cite{QSH}, which have potential applications
in quantum devices. Recently, SO coupling has been induced in ultracold
spinor Bose gases of $^{87}$Rb atoms \cite{SpielmanNature2011} by
the so-called {}``synthetic non-Abelian gauge fields''. Combined
with unprecedented controllability of interactions and geometry in
ultracold atoms, this manipulation of SO coupling opens an entirely
new paradigm for studying strong correlations of quantum many-body
systems under non-Abelian gauge fields.

In this context, over the past few years there have been great theoretical
efforts to determine quantum states of an SO coupled spinor Bose-Einstein
condensate (BEC) \cite{GalitskiPRA2008,WuPreprint,LarsonPRA2009,ZhaiPRL2010,HoPreprint,ZhangPreprint,YouPRA2011}.
In a recent work by Wang \textit{et al.} \cite{ZhaiPRL2010}, two
distinct phases are identified for a {\em homogeneous} two-dimensional
(2D) spin-1/2 BEC. Depending on the relative magnitude of intra-species
($g$) and inter-species ($g_{_{\uparrow\downarrow}}$) interactions,
all bosons can condense into either a single plane-wave state ($g<g_{_{\uparrow\downarrow}}$)
or a density-stripe state ($g>g_{_{\uparrow\downarrow}}$).

The purpose of this Letter is to show that the presence of a {\em
harmonic trap}, which is necessary in experiments, can change dramatically
the phase diagram of SO coupled BECs. At strong SO coupling  the single-particle spectrum 
decomposes into discrete manifolds, analogous to discrete Landau levels.
Non-trivial quantum states with skyrmion lattices emerge when all bosons 
occupy into the lowest manifold (LM). These properties are fundamentally different from that of
a homogeneous system. We note that, in a previous work, the NIST
group has experimentally realized an artificial Abelian gauge field
which leads to the observation of vortex lattice in a non-rotating
$^{87}$Rb condensate \cite{vortex}. Our work represents an important
extention into the regime of non-Abelian gauge field in which the
spin degrees of freedom play an essential role.

\begin{figure}[htp]
\begin{centering}
\includegraphics[clip,width=0.40\textwidth]{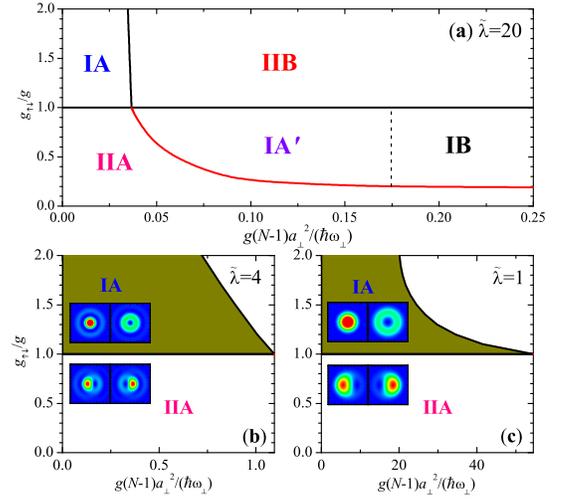} 
\par\end{centering}

\caption{(color online). (\textbf{a}) Phase diagram of a trapped 2D BEC with
a strong SO coupling $\tilde{\lambda}=20$, where the single-particle spectrum forms 
discrete manifolds. For the weak interaction considered here, only the LM 
is occupied. The phases I and II preserve, respectively, the parity and parity-time-reversal
symmetries. There are several sub-phases indicated by A, A$^{\prime}$
and B, which differ in the density profile and/or angular momentum.
The mean-field density patterns in different phases of spin-up bosons
are shown in Fig.~\ref{fig3}. (\textbf{b}) and (\textbf{c}) Phase
diagram at weak SO coupling. Here the phases are determined without
the restriction to the LM approximation. The insets illustrate the density profiles
of the two spin components in phases IA and IIA.}

\label{fig1} 
\end{figure}

Our main results are summarized in Fig. 1, which shows the ground
state as functions of interatomic interaction at a dimensionless SO
coupling strength $\tilde{\lambda}$. By using mean-field theory and exact
diagonalization, we find that: (\textbf{i}) The ground state falls
into two classes of quantum phases, I and II, preserving respectively
the parity (${\cal P}$) and parity-time-reversal (${\cal PT}$) symmetries.
Both symmetries are satisfied by the model Hamiltonian {[}see Eqs.~(\ref{ham})
below{]}. (\textbf{ii}) In each class, there are several sub-phases
(IA, IA$^{\prime}$, IB and IIA, IIB) differing in the density distribution
and/or total angular momentum. (\textbf{iii}) The transition between
different phases depends on interatomic interactions. At weak intra-species
interactions below a critical value, $g<g_{c}$, the ground state
is a half-quantum vortex state (IA) if $g<g_{_{\uparrow\downarrow}}$
and a superposition of two degenerate half-quantum vortex states (IIA)
otherwise. The phases IA and IIA vanish in the limit of strong SO
coupling, but dominate the phase diagram in the opposite. When the
intra-species interactions becomes larger ($g>g_{c}$), there is an
interesting reverse of the symmetry class, i.e., interactions change
the phase IA into IIB and the phase IIA into IA$^{\prime}$ and then
IB. In the phases IIB and IB, skyrmion lattices emerge spontaneously
without rotation. (\textbf{iv}) At $g=g_{_{\uparrow\downarrow}}$,
the phases are ordered by quantum fluctuations. Using exact diagonalization,
we find that the phases follow those at $g<g_{_{\uparrow\downarrow}}$.

\textit{Model Hamiltonian and energy spectrum}. - We consider $N$-bosons
in a 2D harmonic trap $V(\rho)=M\omega_{\perp}^{2}\rho^{2}/2$ with
a Rashba SO coupling ${\cal V}_{so}=-i\lambda_{R}(\partial_{y}\hat{\sigma}_{x}-\partial_{x}\hat{\sigma}_{y})$,
where $\hat{\sigma}_{x,y,z}$ are the Pauli matrices. The model Hamiltonian
is given by ${\cal H=H}_{0}+{\cal H}_{int}$, where \begin{subequations} \label{ham}
\begin{eqnarray}
{\cal H}_{0} & = & \int d{\bf r}\Psi^{+}\left[-\hbar^{2}\nabla^{2}/(2M)+V(\rho)+{\cal V}_{so}\right]\Psi,\label{hami0}\\
{\cal H}_{{\rm int}} & = & \int d{\bf r}\left[(g+g_{\uparrow\downarrow})\hat{n}^{2}+(g-g_{\uparrow\downarrow})\hat{S}_{z}^{2}\right]/4\,,\label{hamiINT}\end{eqnarray}
 \end{subequations} $\Psi=[\Psi_{\uparrow}({\bf r)},\Psi_{\downarrow}({\bf r)}]^{T}$
denotes collectively the spinor Bose field operators, and $\hat{n},\hat{S}_{z}=\Psi_{\uparrow}^{+}\Psi_{\uparrow}{\bf \pm}\Psi_{\downarrow}^{+}\Psi_{\downarrow}$.
We define two characteristic lengths, $a_{\bot}=\sqrt{\hbar/(M\omega_{\perp})}$
for the harmonic trap and $a_{\lambda}=\hbar^{2}/(M\lambda_{R})\ $for
the SO coupling. The dimensionless SO coupling strength can be then
defined as $\tilde{\lambda}=a_{\bot}/a_{\lambda}=(M/\hbar^{3})^{1/2}\lambda_{R}/\omega_{\perp}^{1/2}$.
The Hamiltonian is invariant under two symmetry operations, associated
respectively with the anti-unitary time-reversal operator ${\cal T}=i\sigma_{y}{\cal C}$,
where ${\cal C}$ takes the complex conjugate, and the unitary parity
operator ${\cal P}=\sigma_{z}{\cal I}$, where ${\cal I}$ is the
spatial inversion operator. The Hamiltonian is also invariant under
the combined ${\cal PT}$ operator, which is unitary since ${\cal P}$
and ${\cal T}$ anti-commute with each other, i.e., $\{{\cal P},{\cal T}\}=0$.

In polar coordinates ($\rho,\varphi$), the single-particle eigen-wavefunctions
of ${\cal H}_{0}$ may be written in the form, $\Phi_{m}({\bf r})=[\phi_{\uparrow}(\rho)e^{im\varphi},\phi_{\downarrow}(\rho)e^{i(m+1)\varphi}]^{T}$,
which is energetically degenerate with its time reversed partner ${\cal T}\Phi_{m}({\bf r})=[\phi_{\downarrow}(\rho)e^{-i(m+1)\varphi},-\phi_{\uparrow}(\rho)e^{-im\varphi}]^{T}$.
This degeneracy is a direct consequence of the Kramers' Theorem. Here
we may restrict $m$ to be non-negative integers, as a negative $m$
state can be regarded as the time reversal partner for a state with
$m\ge0$. In this construction, $\Phi_{m}$ and ${\cal T}\Phi_{m}$
are both parity eigenstates with corresponding eigenvalues $(-1)^{m}$
and $(-1)^{m+1}$, respectively. However, they break the ${\cal PT}$
symmetry. The lowest single-particle state occurs at $m=0$ and has
a half-quantum vortex configuration \cite{WuPreprint}. Due to the
degeneracy, any linear superposition of $\Phi_{m}$ and ${\cal T}\Phi_{m}$
--- which breaks the parity symmetry --- is also an eigenstate of
the system. In particular, we may choose the equal-weight superposition
as $(\Phi_{m}+{\cal T}\Phi_{m})/\sqrt{2}$ which can be easily shown
to be eigenstates of ${\cal PT}$.

\begin{figure}[htp]
\begin{centering}
\includegraphics[clip,width=0.45\textwidth]{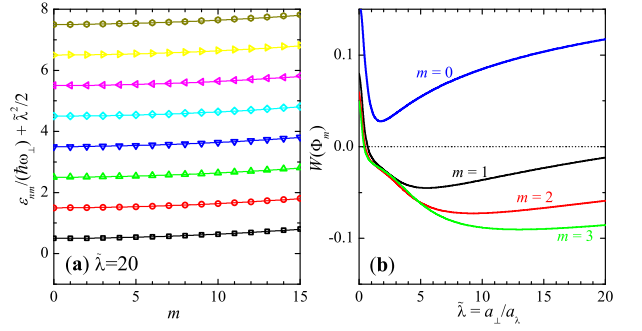} 
\par\end{centering}

\caption{(color online). (a) Single-particle energy spectrum. The lines show
the empirical Eq. (\ref{spectrumSO}). (b) The $W$-function for the
lowest four single-particle states in the LM.}

\label{fig2} 
\end{figure}

The wavefunctions and the corresponding eigenenergies can be found
numerically. At large SO coupling (i.e., $\tilde{\lambda}>5$), to
a good approximation we find numerically that the low-lying spectrum forms discrete manifolds with spacing $\hbar\omega_{\perp}$(indexed
by an integer $n\geq0$), \begin{equation}
\epsilon_{nm}\simeq\left[-{\tilde{\lambda}^{2}}+(2n+{1})+{m\left(m+1\right)}/{\tilde{\lambda}^{2}}\right]\hbar\omega_{\perp}/2\,.\label{spectrumSO}\end{equation}
 There are about $2\sqrt{2}\tilde{\lambda}$ levels within each manifold
with the smallest level spacing $\Delta E=\hbar\omega_{\perp}/\tilde{\lambda}^{2}$.
The discrete manifolds of spectrum are similar to the well-known Landau levels, formed when a charged particle moves in magnetic fields. 
However, the reasons for their formation are very different. In our case of large SO coupling,
without trap the spectrum is characterized by a
continuous momentum ${\bf k}$ and is given by $\epsilon_{{\bf k}}=[-\tilde{\lambda}^{2}/2+(k\pm\tilde{\lambda})^{2}/2]\hbar\omega_{\perp}$,
with infinite degeneracy along the azimuthal direction. The inclusion
of trapping potential quantizes the radial motion for ${\bf k}$ and the
azimuthal motion, giving the standard quantization contribution of
$(n+1/2)\hbar\omega_{\perp}$ and $\left(m+1/2\right)^{2}/(2\tilde{\lambda}^{2})\hbar\omega_{\perp}$
to the energy, respectively.

For a weakly interacting BEC with $gN,\,\, g_{\uparrow\downarrow}N\ll\hbar\omega_{\perp}$,
only the LM is occupied. It is thus convenient to expand the field
operator $\Psi=\sum_{m}\Phi_{m}({\bf r})a_{m}$, where $\Phi_{m}({\bf r})$
is the single-particle wavefunctions at the LM with energy $\epsilon_{m}$.
The many-body Hamiltonian may then be rewritten as, \begin{equation}
{\cal H=}\sum_{m}\epsilon_{m}a_{m}^{+}a_{m}+\sum_{ijkl}V_{ijkl}a_{i}^{+}a_{j}^{+}a_{k}a_{l},\label{hamiLLL}\end{equation}
 where the interaction elements $V_{ijkl}$ can be calculated straightforwardly
for the contact interatomic interactions. We solve Eq. (\ref{hamiLLL})
numerically by using both mean-field theory \cite{ButtsRokhsar} and
exact diagonalization \cite{LiuPRL2001}, for a conserved total angular
momentum $\sum_{m}(m+1/2)a_{m}^{+}a_{m}=Nm_{tot}$. Within mean-field,
we replace $a_{m}$ by a complex number $N^{1/2}c_{m}$ and minimize
the GP energy $E_{GP}/N=\sum_{m}\epsilon_{m}\left|c_{m}\right|^{2}+(N-1)\sum_{ijkl}V_{ijkl}c_{i}^{*}c_{j}^{*}c_{k}c_{l},$
under the constraints $\sum_{m}\left|c_{m}\right|^{2}=1$ and $\sum_{m}(m+1/2)\left|c_{m}\right|^{2}=m_{tot}$.
In practice, we truncate the angular momentum to $\left|m\right|\leq m_{c}$
(up to $m_{c}=16$).

\textit{Symmetry of condensate states}. - In the presence of the interaction
represented by Eq.~(\ref{hamiINT}), the many-body Hamiltonian still
possesses both ${\cal P}$ and ${\cal PT}$ symmetries. As we have
shown above, for a non-interacting system, we may choose the single-particle
ground state to be an eigenstate of ${\cal P}$, or of ${\cal PT}$,
or of neither operator. In the mean-field level, this freedom of choosing
different symmetry eigenstates may be removed by inter-atomic interactions.
In other words, the symmetry of condensate states would be determined
{\em spontaneously} by interaction. We have found that in the weakly
interacting limit we are interested in here, the ground state is either
an eigenstate of ${\cal P}$, or that of ${\cal PT}$. Which symmetry
the ground state will possess can be determined in the following way.
Let us consider an eigenstate of ${\cal P}$ with wavefunction $\Phi_{{\cal P}}=[\phi_{\uparrow}({\bf r}),\phi_{\downarrow}({\bf r})]^{T}$.
The corresponding eigenstate of ${\cal PT}$ can be constructed as
$\Phi_{{\cal PT}}=(\Phi_{{\cal P}}\pm{\cal T}\Phi_{{\cal P}})/\sqrt{2}$.
The mean-field energy difference between these two states is determined
by the $S_{z}^{2}$ term in Eq. (\ref{hamiINT}) which breaks the spin rotational symmetry in the interaction Hamiltonian: \begin{equation}
\Delta E_{sp}\left(\Phi\right)=E(\Phi_{{\cal PT}})-E(\Phi_{{\cal P}})=(g_{\uparrow\downarrow}-g)W\left(\Phi\right)/4,\label{W}\end{equation}
 where $W\left(\Phi\right)\equiv\int d{\bf r[(}\left|\phi_{\uparrow}\right|^{2}-\left|\phi_{\downarrow}\right|^{2})^{2}-(\phi_{\uparrow}\phi_{\downarrow}+\phi_{\uparrow}^{*}\phi_{\downarrow}^{*})^{2}]$.
The ground state will be a ${\cal P}$-eigenstate if $\Delta E_{sp}\left(\Phi\right)>0$
for which we have $n_{\sigma}({\bf r})=n_{\sigma}(-{\bf r})$, or
a ${\cal PT}$-eigenstate if $\Delta E_{sp}\left(\Phi\right)<0$ for
which we have $n_{\uparrow}({\bf r})=n_{\downarrow}(-{\bf r})$. The
$W$-functions of several parity eigenstates are shown in Fig.~\ref{fig2}(b).
Equation (\ref{W}) also shows that the symmetry of the ground state
is sensitive to the relative magnitude of the interaction parameters
$g$ and $g_{\uparrow\downarrow}$.

\begin{figure}[htp]
\begin{centering}
\includegraphics[clip,width=0.40\textwidth]{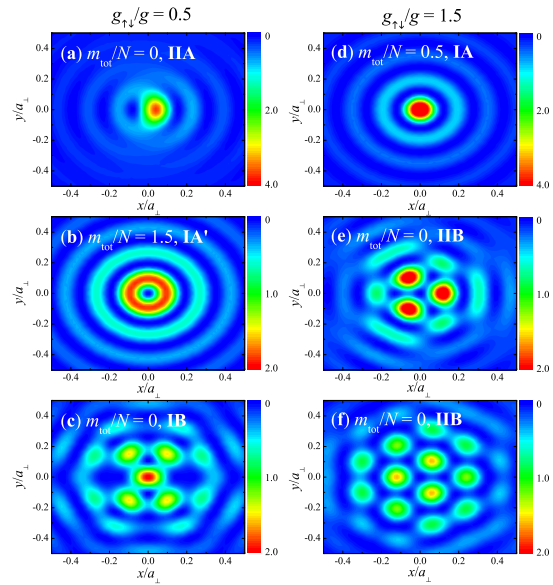} 
\par\end{centering}

\caption{(color online). Density patterns of spin-up bosons in the different
ground states at three intra-species interactions $g(N-1)a_{\perp}^{2}$:
(a,d) $0.02\hbar\omega_{\perp}$, (b,e) $0.1\hbar\omega_{\perp}$,
and (c,f) $0.2\hbar\omega_{\perp}$.}

\label{fig3} 
\end{figure}

\textit{Phase diagram in the LM}. - Our symmetry argument suggests
that all the condensate states could be classified by its ${\cal P}$
or ${\cal PT}$ symmetry, to be referred to respectively as phases
I and II hereafter. We now check numerically this argument in the
quantum Hall like regime with all bosons occupying into the LM, as
shown in Fig.~\ref{fig1}(a) for $\tilde{\lambda}=20$. The characteristic
density distributions for spin-up bosons in each phase are shown in
Fig.~\ref{fig3}.

At sufficiently weak interactions, where the characteristic interaction
energy $g(N-1)a_{\perp}^{2}$ is smaller than the lowest intra-manifold
spacing $\Delta E=\hbar\omega_{\perp}/\tilde{\lambda}^{2}$, only
the ground single-particle state is occupied. The condensate state
is thus either half-quantum vortex states of $\Phi_{0}$ (or ${\cal T}\Phi_{0}$)
or their superposition. As $W(\Phi_{0})>0$ as shown in Fig.~\ref{fig2}(b),
we conclude that the ground state is a ${\cal PT}$-eigenstate for
$g>g_{\uparrow\downarrow}$ (IIA) and it is a half-quantum vortex
state (a ${\cal P}$-eigenstate) for $g<g_{\uparrow\downarrow}$ (IA).
Their spin-up density patterns are shown in Figs.~\ref{fig3}(a)
and (d), respectively.

\begin{figure}[htp]
\begin{centering}
\includegraphics[clip,width=0.40\textwidth]{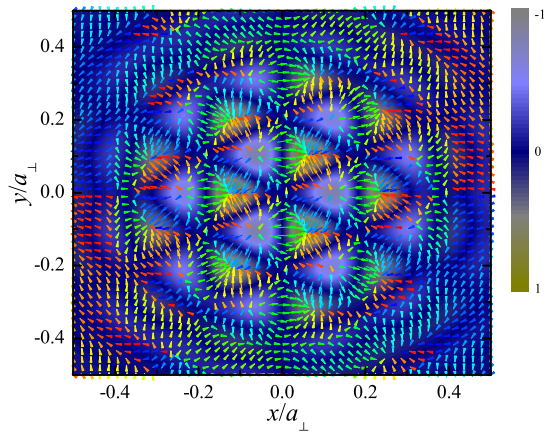} 
\par\end{centering}

\caption{(color online). Spin texture $\mathbf{S}=(1/2)\Phi^{+}\mathbf{\mathbf{\sigma}}\Phi$
corresponding to the state represented in Fig.~\ref{fig3}(f). The
arrows represent the transverse spin vector ($S_{x},S_{y}$) with
color and length representing the orientation and the magnitude of
the transverse spin. The contour plot shows the axial spin $S_{z}=(1/2)\left(\phi_{\uparrow}^{2}-\phi_{\downarrow}^{2}\right)$.}

\label{fig4} 
\end{figure}

When the interaction becomes larger, more and more single-particle
states are occupied. The occupation of the first excited single-particle
state ($\Phi_{1}$ and ${\cal T}\Phi_{1}$) occurs at $g_{c}(N-1)a_{\perp}^{2}\simeq0.0367\hbar\omega_{\perp}$,
where the critical interaction strength $g_{c}$ is determined from
the equation $\epsilon_{0}+(N-1)V_{0000}=\epsilon_{1}+(N-1)V_{1111}$.
As $W(\Phi_{m})<0$ for $m\geq1$, we find an interesting reverse
of the phase diagram when $g>g_{c}$: the ${\cal P}$-preserving phase
(IA) changes into a ${\cal PT}$-preserving phase (IIB) at $g<g_{\uparrow\downarrow}$,
while the ${\cal PT}$-preserving phase (IIA) changes into a ${\cal P}$-preserving
phase (IA$^{\prime}$ and IB) if $g>0.2g_{\uparrow\downarrow}$. The
phases IA$^{\prime}$ and IB differ in the total angular momentum
$m_{tot}$ and density distribution. In Phase IB, $m_{tot}$ is suppressed
to zero by large interatomic interactions. Note that in the phases
(IIB) and (IB), we observe regular lattice patterns. In particular,
a hexagonal lattice form gradually in the phase IIB, as shown clearly
in Figs.~\ref{fig3}(e) and (f). In Fig.~\ref{fig4}, we show the
corresponding spin texture of the state, from which one can see that
the system represents a lattice of skyrmions. Skyrmion lattice can
be generated by rotating a spinor condensate \cite{guo}. Here 
the skyrmion texture is induced by the SO coupling without rotation.

The symmetry of the ground state at $g=g_{\uparrow\downarrow}$ can
not be determined within mean-field theory, since in this case $\Delta E_{sp}\left(\Phi\right)=0$
{[}see Eq.~(\ref{W}){]} and the energy becomes invariant for different
$m_{tot}$. However, it can be ordered by quantum fluctuations \cite{WuPreprint},
which are well captured by exact diagonalization. We have calculated
the energy as a function of $m_{tot}$ at $g(N-1)a_{\perp}^{2}/(\hbar\omega_{\perp})=0.02$
and $0.1$ for $N=4$, $8$, and $12$. With increasing $N$, the
exact diagonalization result approaches the mean-field prediction.
We find that the ground state at $g(N-1)a_{\perp}^{2}=0.02\hbar\omega_{\perp}$
has a spontaneous angular momentum $m_{tot}=-1/2$ or $+1/2$, while
the ground state at $g(N-1)a_{\perp}^{2}=0.1\hbar\omega_{\perp}$
occurs at $m_{tot}=0$. Therefore, we identify that the phases at
$g=g_{\uparrow\downarrow}$ follow those at $g<g_{\uparrow\downarrow}$.
This is in agreement with the result of Ref.~\cite{WuPreprint},
which employs a different {}``order from disorder'' argument.

\textit{Phase diagram beyond LM}. - So far we have clarified the phase diagram
at a particular SO coupling $\tilde{\lambda}=20$ in the \emph{weakly-interacting}
LM regime. However, the qualitative picture of diagram may persist
beyond the regime of LM, as far as our symmetry argument holds. To check this, we performed a direct numerical calculation based on the full Gross-Pitaevskii (GP) equation derived from Eqs.~(\ref{ham}) without making the LM assumption. In the regime as shown in Fig.~\ref{fig1}(a), the results are in good agreement with the LM calculation. At larger interaction strength when higher manifolds get mixed in the ground state, we have found from the GP calculation that Phase IIB in Fig.~\ref{fig1}(a) will change to a density-stripe phase with ${\cal P}$
symmetry, while Phase IB will change to a plane-wave phase with ${\cal PT}$ symmetry. The density-stripe and the plane-wave phases have been shown to be the mean-field ground state for a {\em homogeneous} system \cite{ZhaiPRL2010}. For the trapped system as studied here, at large interaction strength, the effect of the trap becomes less important and our results are therefore consistent with those reported in Ref.~\cite{ZhaiPRL2010}.  
With decreasing $\tilde{\lambda}$, we anticipate that the phases
IA and IIA will gradually become dominant in the diagram, as we find
numerically that $g_{c}\propto 1/\tilde{\lambda}^{2}$ increases very
rapidly. The skyrmion lattice phase, related
to the LM formation, may disappear. This is confirmed by the GP calculation for smaller SO coupling
and the results are represented in Fig.~\ref{fig1} (b) and (c). The half-quantum vortex state and its superposition dominate over a much larger parameter space as compared to the large SO coupling case. A more detailed study of the complete phase diagram and the properties of different phases will be presented elsewhere \cite{RamPreprint}.

\textit{Experimental relevance}. - We finally consider the 
experimental feasibility. A Rashba SO coupling can be induced in spinor $^{87}$Rb
gases \cite{SpielmanNature2011}. The interaction strengths of $^{87}$Rb
atoms may be tuned by properly choosing the parameters of the laser
fields that induce the SO coupling \cite{ZhangPreprint}. The two-dimensionality
in such system has now been routinely realized by imposing a strong
harmonic confinement $V(z)=M\omega_{z}^{2}z^{2}/2$ along the $z$-direction with $\omega_z \gg \omega_\perp$.
The
critical temperature for an ideal 2D SO BEC is given by $T_{c}=(c_{\lambda}/\pi)\sqrt{3N} \hbar\omega_{\perp}/k_{B}$,
where the prefactor $c_{\lambda}<1$ takes into accout the suppression
due to the SO coupling. Taking parameters from a recent experiment \cite{Dalibard2D} with $\omega_{\perp}=2\pi\times20.6$ Hz and $N\sim 10^{5}$, we find at $\tilde{\lambda}=10$, $c_{\lambda}\sim 0.6$
and $k_{B}T_{c}\simeq 120$ nK. Experimentally, BEC temperature below 0.5 nK has been recorded \cite{StockLPL2005}, which is also lower than $\hbar \omega_\perp/k_B$. The mean-field LLL regime is therefore readily attainable with current technologies.

\textit{Conclusion}. - In summary, we have investigated the phase
diagram of a spin-orbit coupled spinor BEC in harmonic traps, by using
mean-field theory and exact diagonalization method. We have predicted
that the condensate states preserve the parity or parity-time-reversal
symmetry and exhibit spontaneous vortex and skyrmion lattice structure
in the lowest energy manifold which is induced by strong spin-orbit coupling.
Our results are valid for weak correlations with large number of bosons.
Strongly correlated states, analogous to the fractional quantum Hall
states, would emerge with small number of bosons \cite{WilkinGunn}.
These can be addressed using exact diagonalization method in future
studies.

\textit{Acknowledgment} --- We would like to thank Hui Zhai, Congjun
Wu, Xiang-Fa Zhou and Shih-Chuan Gou for useful discussions. HH and
XJL were supported by the ARC Discovery Projects (Grant Nos. DP0984522
and DP0984637) and NFRP-China (Grant No. 2011CB921502). HP was supported
by the NSF, the Welch Foundation (Grant No. C-1669) and the DARPA
OLE program.

\textbf{Note added.} - When our manuscript was under review, we became aware of a preprint \cite{SinhaPreprint}, in which the authors
addressed the same problem at $g=g_{\uparrow\downarrow}$.

\end{document}